\documentclass[aps,groupedaddress,10pt,twocolumn,amsmath,amssymb]{revtex4-1}
\usepackage{epsfig}
\usepackage{xspace}
\usepackage{amssymb,amsmath}
\usepackage{graphicx}

\begin{document}

\title{Hydrogen bonds and van der Waals forces in ice at ambient and high pressures}

\author{Biswajit Santra$^1$}
\author{Ji\v{r}\'{i} Klime\v{s}$^{2,3,4}$}
\author{Dario Alf\`e$^{2,4,5,6}$}
\author{Alexandre Tkatchenko$^1$}
\author{Ben Slater$^{3,4}$}
\author{Angelos Michaelides$^{2,3,4}$}
\email{angelos.michaelides@ucl.ac.uk}
\author{Roberto Car$^{7}$}
\author{Matthias Scheffler$^1$}
\affiliation{$^1$Fritz-Haber-Institut der Max-Planck-Gesellschaft, Faradayweg 4-6, 14195 Berlin, Germany \\
$^2$London Centre for Nanotechnology, University College London, London WC1E
6BT, UK \\
$^3$Department of Chemistry, University College London, London WC1E 6BT, UK \\
$^4$Thomas Young Centre, University College London, London WC1E 6BT, UK \\
$^5$Department of Physics and Astronomy, University College London, London WC1E 6BT, UK \\
$^6$Department of Earth Sciences, University College London, London WC1E 6BT, UK \\
$^7$Department of Chemistry, Princeton University, Princeton, New Jersey 08544, USA}

\begin{abstract}

The first principles approaches, density functional theory (DFT) and quantum Monte Carlo, 
 have been used to examine the balance between van der Waals
(vdW) forces and hydrogen (H) bonding in ambient and high pressure phases of ice.
At higher pressure, the contribution
to the lattice energy from vdW increases and that from H bonding decreases, 
leading vdW to have a substantial effect on the transition pressures
between the crystalline ice phases.
An important consequence, likely to be of relevance to molecular crystals in general, is that transition pressures 
obtained from DFT functionals which neglect vdW forces are greatly overestimated.
\end{abstract}

\maketitle
\pagebreak

Water-ice, the most common molecular solid in nature, exhibits a rich and complex phase diagram.
At present this includes ice Ih, ice Ic and 14 other crystalline ice phases \cite{salzmann_09,mishima_85}.
Although the phase diagram of ice up to around 2 GPa is well-established experimentally, understanding 
of the subtle balance of intermolecular interactions which give rise to this richness is incomplete.
In particular, the relative contribution of hydrogen (H) bonding and van der Waals (vdW) dispersion forces 
to the cohesive properties of the various crystalline ice phases is still not understood. 
From a theoretical perspective, the varied densities of experimentally characterized 
ice crystal structures, with the molecules fixed on well-defined lattices, affords an excellent opportunity to quantify and assess
the nature of these intermolecular interactions.

Computer simulation techniques have proved instrumental in understanding ice (e.g. \cite{car_ice_92,hamann_97,vega_PRL_04,klein_05,slater_jacs_06,de_koning_06,feibelman_08,hermann_08,pan_08,pisani_09,car_09,militzer_10}).
%
%
In particular, density-functional theory (DFT) with 
generalized gradient approximation (GGA) 
functionals
%
has been widely applied.
Certain GGAs describe the ambient pressure ice Ih phase reasonably well \cite{feibelman_08}
and predict the proton order--disorder phase transition temperatures between ice Ih and XI and ice VII and VIII
in good agreement with experiments \cite{klein_05}.
%
%
However, it is known that GGAs suffer deficiencies when vdW
forces are important and indeed it has been suggested that this is
the reason certain GGAs produce a liquid water density about 15--20\% less
than experiment \cite{lin_jpcB_09,schmidt_jpcB_09,artacho_10} or fail to predict the correct ground state structure for
some small water clusters \cite{santra_hexamer}.
Despite the considerable body of DFT work on ice phases, the role of vdW
forces and H bonding has not been systematically examined in anything other 
than ice Ih \cite{hamada_10,labat_11}.
%
However, with recent developments (e.g. \cite{Dion_vdw_04,TS_vdw_09,grimme_vdw_10}),
it is now possible to tackle this issue head on and estimate the importance
of vdW forces and H bonding in the various phases of ice.
%


%
Here we report an extensive series of first principles studies aimed at better understanding
the role of vdW and H bonding in ice.
This includes DFT calculations with and without a treatment of 
vdW forces 
on the ambient pressure phase of ice, ice Ih, 
and all the proton ordered phases, namely,
in order of increasing pressure,
ice IX, II, XIII, XIV, XV, VIII.
We also report diffusion quantum Monte Carlo (DMC) 
calculations
for ice Ih, II, and VIII
as reference values to complement the experimental results.
DMC is highly accurate for weak interactions
including vdW bonded systems 
and represents the state-of-the-art for electronic structure simulations of solids.
From this work we find that the contribution of vdW to the
cohesive properties of ice increases as one moves to the higher density phases, 
whereas the contribution from H bonds decreases.
As a result,
vdW plays a crucial role in determining the relative
stabilities and phase transition pressures in ice.
The results presented here are likely to be of relevance to understanding intermolecular interactions in water in all its condensed phases as well as to structural searches for (novel) high pressure ices and to molecular crystals in general.

\begin{figure*}
   {\includegraphics[width=14cm]{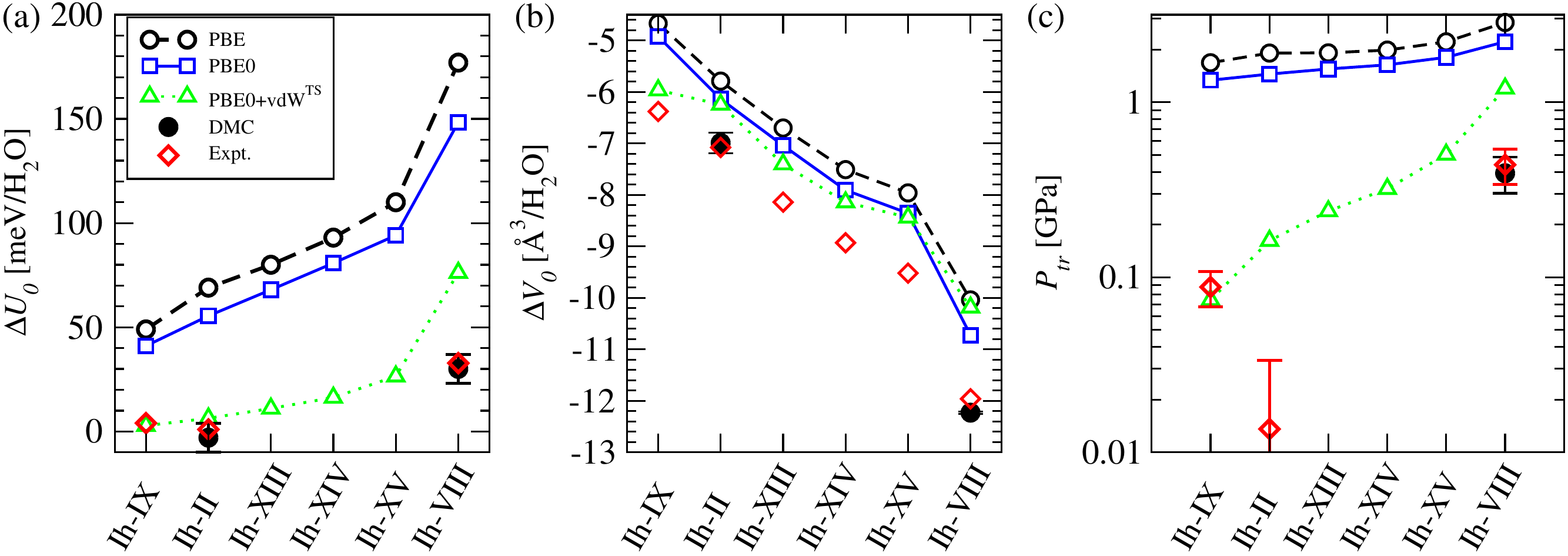}}
\caption{\label{pressure} Relative lattice energies, $\Delta U_0$ (a), and volumes, $\Delta V_0$ (b), of the high
pressure ice phases with respect to the lattice energy of ice Ih obtained with the methods indicated.
(c) Transition pressures ($P_{tr}$) from ice Ih to the various high pressure phases.}
\end{figure*}

\begin{table}
\caption{\label{tb1} Absolute lattice energies (omitting zero point energy effects) of ice Ih, II, and VIII. Relative energies compared to ice Ih
($\Delta U_0$) are given in parenthesis. All values are in meV/H$_2$O.}
\begin{ruledtabular}
\begin{tabular}{c|ccc}
	      & Ih      & II         & VIII    \\
\hline
Expt.\footnote{Ref. \cite{whalley_jcp_84}, with zero point energy contributions removed.}
        & -610    &   -609 (1)         &     -577 (33)                \\
DMC       & -605 $\pm$5  & -609 $\pm$5 (-4)  & -575 $\pm$5 (30)  \\
PBE       & -636   & -567 (69) & -459 (177)     \\
PBE0      & -598    & -543 (55) & -450 (148) \\
PBE0+vdW$^{\rm TS}$    & -672     & -666 (6)     & -596 (76) \\
%
\end{tabular}
\end{ruledtabular}
\end{table}

%
%

We start by discussing lattice energies,
which are obtained by subtracting the total energy  \cite{supplement} of the ice unit cell containing $n$ H$_2$O molecules
from the total energy of $n$ isolated H$_2$O molecules.
In this context Whalley's extrapolations of the experimental finite temperature and pressure phase coexistence
lines to zero temperature and pressure are extremely valuable \cite{whalley_jcp_84}.
For the proton ordered phases Whalley considered, these indicate that ice IX, II, and VIII are less stable
than ice Ih by only 3.5 $\pm$ 0.8, 0.6 $\pm$ 1.0, and 33 meV/H$_2$O, respectively \cite{whalley_jcp_84}.
These values agree well with DMC.
Specifically, DMC predicts that ice VIII is 30 $\pm$ 7 meV/H$_2$O less stable than ice Ih.
Similarly, the near energetic degeneracy between ice Ih and ice II is
captured with DMC, and within the DMC error bars, ice Ih and ice II are equally stable.
From Table I it can also be seen that DMC lattice energies are in very good agreement
with experiment as well.
Overall this gives us confidence in the quality of
Whalley's experimental extrapolations and
the accuracy of the DMC calculations.
This therefore provides an excellent basis for exploring the role of
H bonds and vdW in ice with  DFT.


We now discuss the results obtained with PBE \cite{PBE}, one of the most
widely used functionals.
%
For ice Ih, PBE yields a reasonable lattice energy of about 640 meV/H$_2$O,
an overestimate of around 30 meV/H$_2$O
which is consistent with previous work \cite{feibelman_08}.
However, Table I and Fig. 1(a) reveals a severe deterioration in
the performance of PBE for the higher density phases.
Whereas experiment and DMC suggest that the energy difference between ice Ih
and the least stable ice VIII phase is about 30 meV/H$_2$O, PBE gives $\sim$180 meV/H$_2$O difference.
%
This is mainly because PBE underestimates the stability of the
high pressure phases,
suggesting that attractive interactions, more important in the higher density ice phases,
are not captured accurately with PBE.
In addition to substantially overestimating the energy differences between the various phases,
the changes in volumes ($\Delta V_0$) upon going from ice Ih to the high pressure phases
are also underestimated with PBE (Fig. 1(b)) and critically the transition
pressures ($P_{tr}$) --- obtained from $P_{tr} = -\Delta U_0/\Delta V_0$
--- are $\sim$5-15 times larger than experiment (Fig. 1(c)).


Seeking to understand why PBE performs so poorly for the high pressure ice phases, we
have considered various potential sources of errors. 
First we analyzed the dipole moment and polarizability of an isolated water molecule. 
PBE, like other GGAs, overestimates the polarizability of an isolated water molecule by $\sim$10\% compared to experiment, 
which is related to any overly delocalized electron density and a too small HOMO-LUMO gap. 
Incorporation of a fraction of Hartree-Fock exchange 
in to the GGAs is an established approach 
for alleviating this problem.
In particular, a popular PBE-based hybrid functional, PBE0 \cite{PBE0}, 
widens the HOMO-LUMO gap of the isolated water molecule by $\sim$40\% and provides 
more accurate polarizabilities and interaction energies within a variety of 
water clusters \cite{hammond_09,sms,jordan_10}.
When applied to ice, 
the lattice energies obtained with PBE0 
(Table I) are indeed improved over PBE, 
however, the energy difference between ice Ih and ice VIII is
still about four times larger than the experimental value.
In addition, as with PBE, upon going from ice Ih to the high pressure phases the transition 
pressures are about an order of magnitude too large compared to experiment (Fig. 1(c)).
%


%
Turning to the role of H bonding, the
relative H bond strength has been estimated 
from shifts in the O-H stretching frequencies, which is a simple and widely used measure of H bond strength (e.g. \cite{Li_PNAS_2011}).
Specifically, the softening (red-shift) of the intramolecular O-H stretching 
frequencies in the various phases is compared to the (average) O-H stretching frequency of an isolated water monomer \cite{note_on_HB}.
Based on the available experimental frequencies for several of the phases \cite{whalley_64,bertie_77,whalley_76}, 
H bonds get weaker with increasing pressure. 
This established, but not necessarily obvious, result arises because the nearest neighbor water-water  
distances get larger as one moves from ice Ih with its open ring structure to the more complex higher pressure phases. 
In line with previous calculations \cite{cpl_10}, PBE reproduces the trend in frequency changes with pressure but the red-shift (i.e., strength) compared to the the isolated water molecule is significantly 
overestimated compared to experiment, particularly for ice Ih (see Fig. 2(a)).
PBE0 predicts frequency shifts in much better agreement with experiment (Fig. 2(a)). 
However, as we have seen, the relative
stabilities of the ice phases with PBE0 are significantly in error, which clearly suggests that
H bonds are not the only important interaction in the high density phases.


We now consider the influence of vdW interactions with the
scheme of Tkatchenko and Scheffler (referred to as vdW$^{\rm TS}$), 
in which an additional $C_6/R^6$ tail is added to the DFT total energy, 
with the $C_6$ coefficients calculated as functionals of the electron density \cite{TS_vdw_09}.
Dramatic improvements in the relative energies of
the various phases and phase transition pressures are observed (Fig. 1).
%
In particular, upon going from PBE0 
to PBE0+vdW$^{\rm TS}$ 
the energy difference between ice 
Ih and ice II is reduced to just 6 meV/H$_2$O 
and, likewise, ice VIII is now only 76 meV/H$_2$O 
less stable than ice Ih. 
%
As a consequence, with vdW the ice Ih to ice II transition pressure is reduced by more than one order 
of magnitude and is in good agreement with experiment. 
%
Similarly, the 
ice Ih to ice VIII transition comes within a factor of two of experiment.
The substantially improved transition properties when vdW is accounted for results from a strong 
dependence of vdW on density, as shown in Fig. 2(b).
%
Clearly as the density of the ice phases increases, so too does the
vdW contribution to the lattice energy and indeed for the highest density phase (ice VIII) the
vdW contribution is about twice what it is in ice Ih. 
Overall the decrease of H bond strength and the increase of vdW
in the high pressure phases clearly shows a significant enhancement of
relative contribution of vdW over H bonding interactions
for the cohesive properties of the high pressure ice phases.
%


%
Our findings regarding the importance of vdW
for the high-pressure phases of ice also hold when using functionals based on the
vdW-DF approach of Dion \emph{et al.}~\cite{Dion_vdw_04,klimes_10}.
The phase transition pressures obtained with such functionals are similar to those obtained with the TS scheme.
Both approaches neglect the non-additive many-body vdW energy beyond
the pairwise approximation. 
We have found that the many-body vdW energy
plays a minor role for the different phases of ice by using
an extension of the TS scheme~\cite{tkatchenko_11}.
Also we find that quantum effects based on zero point energies play a minor role in 
determining the relative stabilities of the various phases.
Specifically, the zero point energies of the different phases (calculated with PBE and the harmonic approximation) differ by $<$10 meV/H$_2$O.

\begin{figure}
   \includegraphics[width=8cm]{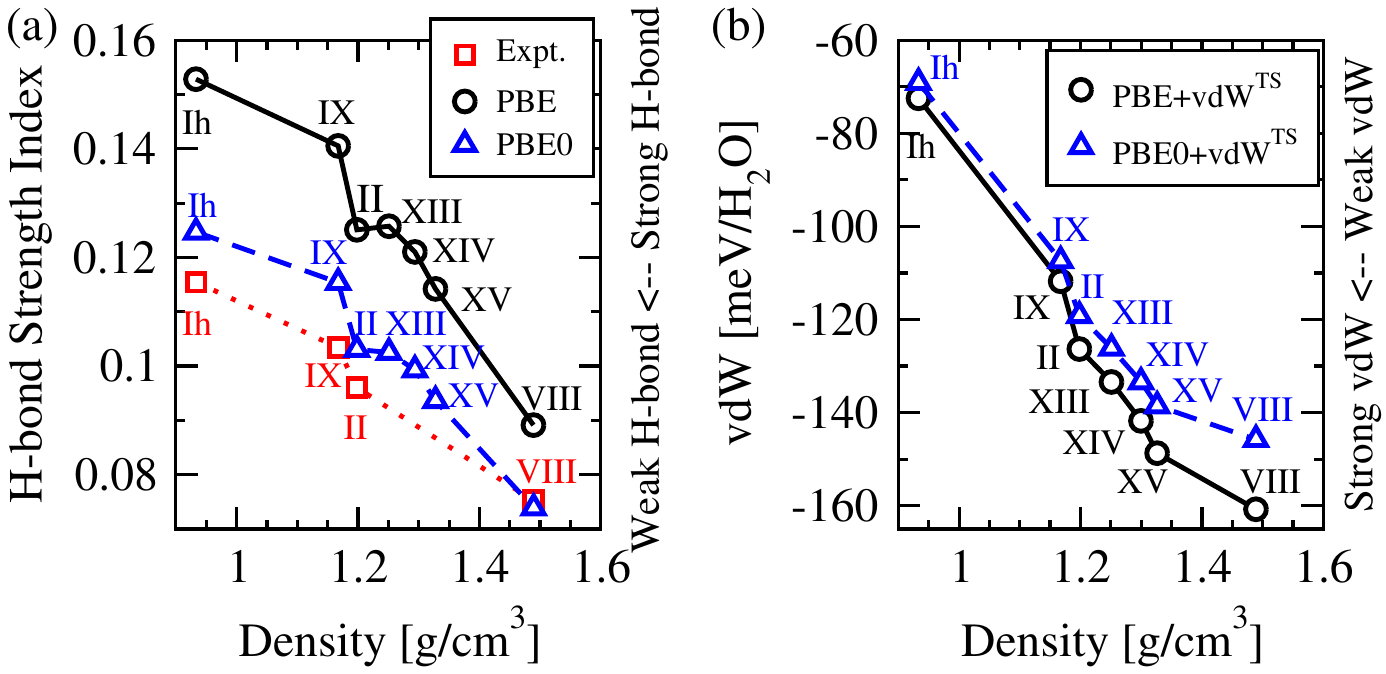}
\caption{\label{hb_vdw} 
(a) H bond strength index \cite{note_on_HB} and (b) vdW energy contributions 
plotted as a function of the experimental densities of the ice phases 
at zero pressure.
Experimental values are taken from \cite{whalley_64,bertie_77,whalley_76}.
%
%
}
\end{figure}


We now discuss why vdW interactions play a crucial role in determining the relative stabilities of
the various ice phases.
For any two non-bonded atoms vdW forces result from an induced dipole-induced dipole
interaction whose leading term varies as $C_6R^{-6}$.
At a given $R$, vdW will increase if the polarizability of the atoms increases, i.e., $C_6$ becomes larger
and/or the number density of the atoms increases.
%
Our calculations with the TS approach show that the $C_6$ coefficients
do indeed increase upon going from ice Ih to ice VIII, by 24\% for H and 6\% for O.
%
%
%
%
However, the major effect that leads to an increase in the vdW interactions in the high density phases is
simply the higher packing of water molecules.
%
This can be seen by comparing the O-O radial distribution functions of
ice Ih with e.g. ice II and ice VIII (Fig. 3(a)).
Although ice Ih possesses the shortest nearest neighbor O-O distances, the structure
is open and there is a large gap of $\sim$2 \AA\ between the first and second coordination shells.
In contrast, in the high density phases the second and subsequent coordination shells appear at much
shorter O-O separations.  
In fact in ice VIII, which is comprised of two interpenetrating sub-lattices, the first and second coordination shells
fall almost on top of each other (with the shortest O-O distances associated with non H bonded contacts).
The higher packing, particularly in the $ca.$ 3 to 6~\AA\ regime is reflected by the
integrated number of neighbors versus O-O distance shown in Fig. 3(b)). 
Overall these additional molecules at short (ice VIII) and intermediate (ice II) distances 
lead to enhanced vdW interactions in the high pressure phases.
Fig. 3(c) provides a more quantitative basis for the above argument,
by showing the total vdW interactions --- as obtained from the TS scheme --- for ice Ih, II and VIII in
the range of $ca.$ 3~\AA\  to 6~\AA.
It can be seen that beyond about $ca.$ 3~\AA\ there are more substantial contributions
from vdW from ice II and VIII than ice Ih.
Although vdW is generally considered to be a long range interaction, the dominance of vdW
in this $ca.$ 3~\AA\  to 6~\AA\  regime is to be expected given that the sum of the vdW radii  of
O and H in the condensed phase (calculated with TS scheme) is only $\le$3~\AA.
Note that in fact the difference in the vdW contribution between the various phases is essentially converged
to the periodic limit after about 8~\AA.

\begin{figure}
   \includegraphics[width=8cm]{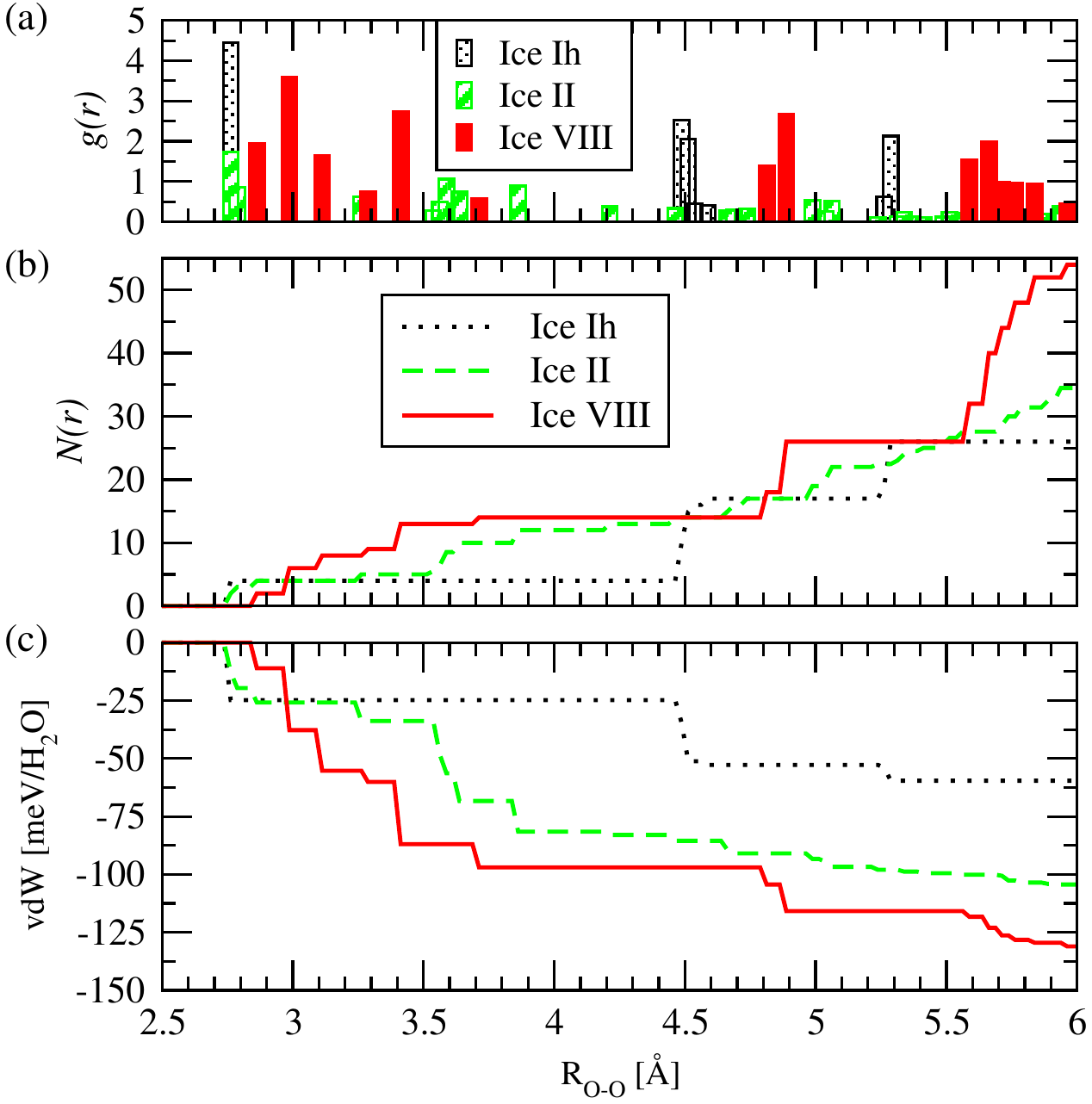}
\caption{\label{vdw2} (a) O-O radial distribution functions ($g(r)$) of ice Ih, ice II, and ice VIII. (b) Integrated number of neighbors ($N(r)$)
and (c) vdW contributions (obtained from the TS scheme) as a function of O-O distance for the same three phases.}
\end{figure}

%
In conclusion, we have performed an extensive first principles study of ice
at ambient and high pressures.
From the ambient to the high pressure phases the contribution
to the lattice energy arising from vdW forces monotonously increases.
This has a significant impact on the phase transition pressures, as exemplified by calculations with
conventional \emph{xc} functionals where vdW forces are not accounted for, and where transition
pressures more than one order of magnitude larger than experiment are obtained.
By accounting for
vdW forces, 
the phase transition pressures more closely agree with experiment.
This finding provides a new physical insight and is also relevant to DFT-based structural
searches for novel ice polymorphs \cite{pickard_2007,militzer_10}, by implying that
a reexamination of the high pressure region of the ice phase diagram with a vdW corrected 
DFT approach would be worthwhile.
Whilst the focus of the current study has been on understanding the role of
H bonding and vdW forces in ice, rather than on improving the current state-of-the-art in DFT simulations 
for water, it is nonetheless worthwhile to stress that there certainly remains scope for
further improvements within DFT in terms of transition pressures, absolute lattice energies 
and volumes for ice.
In particular, there is scope for improving the lattice energy of ice VIII obtained with 
vdW$^{TS}$, with the current underestimation in the lattice energy possibly associated
with the radically different structure of ice VIII compared to the other phases considered.
%
%
The strong dependence of vdW interactions on density suggests that
even at ambient pressure, vdW forces will also be important to liquid water which has a $\sim$8\%
higher density than ice Ih.
Indeed this is consistent with a number of recent DFT studies \cite{schmidt_jpcB_09, artacho_10}.
Although vdW is often associated with so-called
``sparse'' matter we have shown that it attains greater significance at high density.
This somewhat counterintuitive result arises because vdW forces mainly enhance interactions between
molecules at medium range (i.e., second and third coordination shells at 3 to 6 \AA\ separations).
Therefore, analogous effects are expected for water in other environments such as in confined geometries, 
interfaces, and clathrates, and indeed for other hydrogen bonded molecular crystals at high pressure.

%
{\bf Acknowledgements:} 
The work of A.M. and D.A. is supported by a EURYI award.
A.M. is also supported by the EPSRC, and the European Research Council.
J.K is grateful to UCL and the EPSRC for support through 
the PhD+ scheme.
RC is supported by NSF CHE-0956500 and by a senior scientist Award of the Alexander von Humboldt Foundation.  
This research used resources of the Oak Ridge Leadership Computing Facility,
located in the National Center for Computational Sciences at Oak Ridge National
Laboratory, which is supported by the Office of Science of the Department of
Energy under Contract DE-AC05-00OR22725.
We are also grateful for computational resources to the London Centre for Nanotechnology
and UCL Research Computing as well as to the UK's HPC Materials Chemistry Consortium, which is
funded by EPSRC (EP/F067496), for access to 
HECToR, the UK's national high-performance computing service.

%


\begin{thebibliography}{10}
\bibitem{note_on_HB}%
  \BibitemOpen
  \bibinfo {note} {The H bond strength index is calculated as $(\nu_{\rm
  O-H}^{\rm H_2O}-\nu_{\rm O-H}^{\rm ice})/\nu_{\rm O-H}^{\rm H_2O}$, where
  $\nu_{\rm O-H}^{\rm H_2O}$ and $\nu_{\rm O-H}^{\rm ice}$ are the average
  uncoupled covalent O-H stretching frequencies of an isolated H$_2$O molecule
  and ice, respectively. When there is no H bond the index is zero. Frequency
  shifts are just one measure of relative H bond strength. Other key quantities
  related to HB strength, e.g., water molecule dipole moments and covalent O-H
  bond lengths, also decrease from the low to the high density phases.}%
  \bibAnnoteFile{Stop}{note_on_HB}%
\end{thebibliography}

\end{document}